\documentstyle[12pt]{article}

\newcounter{sectionc}\newcounter{subsectionc}\newcounter{subsubsectionc}
\renewcommand{\section}[1] {\vspace{12pt}\addtocounter{sectionc}{1}
\setcounter{subsectionc}{0}\setcounter{subsubsectionc}{0}\noindent
        {\bf\thesectionc. #1}\par\vspace{5pt}}
\renewcommand{\subsection}[1] {\vspace{12pt}\addtocounter{subsectionc}{1}
        \setcounter{subsubsectionc}{0}\noindent
        {\bf\thesectionc.\thesubsectionc. {\kern1pt \bf\it
#1}}\par\vspace{5pt}}
\renewcommand{\subsubsection}[1] {\vspace{12pt}\addtocounter{subsubsectionc}{1}
        \noindent{\thesectionc.\thesubsectionc.\thesubsubsectionc.
        {\kern1pt \it #1}}\par\vspace{5pt}}

\def\thebibliography#1{\leftline{\large \it References}\list
  {[\arabic{enumi}]}{\settowidth\labelwidth{[#1]}\leftmargin\labelwidth
    \advance\leftmargin\labelsep
    \usecounter{enumi}}
    \def\newblock{\hskip .11em plus .33em minus .07em}
    \sloppy\clubpenalty4000\widowpenalty4000}

\newcommand{\ie}{{\it i.e.}\ }
\newcommand{\eg}{{\it e.g.}\ }
\newcommand{\etc}{{\it etc.}\ }

\renewcommand{\sc}{{\it s.c.}}
\newcommand{\be}{\begin{eqnarray}}
\newcommand{\ee}{\end{eqnarray}}
\newcommand{\dslash}{\partial \!\!\! /}

\newcommand{\TTr}{{\bf Tr}}
\newcommand{\tr}{{\rm tr}}

\newcommand{\A}{{\cal A}}

\newcommand{\G}{{\cal G}}
\newcommand{\GG}{{\bf G}}
\renewcommand{\L}{{\cal L}}

\begin{document}

\rightline{UNIT\"U-THEP-4/1995}
\rightline{TU-GK-95-003}
\rightline{hep-ph/jjmmnnn}
\rightline{February 1995}
\vskip .5truecm
%\rightline{\it PACS: 11.10.LM, 14.20.-c}
\vskip 2truecm
\centerline{\Large\bf Diquarks in a chiral soliton field$^\dagger$}
\baselineskip=18 true pt
\vskip 2cm
\centerline{U.\ Z\"uckert$^{1,2}$, R.\ Alkofer,
H.\ Weigel$^3$, and H.\ Reinhardt}
\vskip .3cm
\centerline{Institute for Theoretical Physics}
\centerline{T\"ubingen University}
\centerline{Auf der Morgenstelle 14}
\centerline{D-72076 T\"ubingen, Germany}
\vskip 3cm
\centerline{\bf ABSTRACT}
\vskip .25cm

In the Nambu--Jona-Lasinio model baryons are either considered
as quark-diquark-composites or the soliton configurations of the model
are interpreted as baryons. As a first step towards constructing
a hybrid model, which possesses a dynamical interplay between both
pictures, the Bethe-Salpeter equation for diquarks in the background of
a soliton configuration is solved. The presence of the soliton
causes a significant reduction of the resulting bound state
eigenenergy of the diquark. As a consequence a unit baryon number
configuration may
be constructed with a mass lower than the energy of the soliton. An
estimate of the influence of the diquark on the meson fields indicates
a small decrease of the extension of the soliton due to diquark
correlations.

\vfill
\small
\noindent
------------------

\noindent
$^\dagger$ Supported by COSY under contract 41170833.\hfil\break
$^1$ Member of the Graduiertenkolleg ``Struktur und Wechselwirkung
von Hadronen und Kernen'' of the Deutsche Forschungsgemeinschaft (DFG
Mu 705/3).\hfil\break
$^2$ E-mail: zueck@ptdec1.tphys.physik.uni-tuebingen.de\hfil\break
$^3$ Supported by a Habilitanden scholarship of the DFG.
\eject

\normalsize
\baselineskip=18 true pt

\section{\large \it Introduction}

Two distinct approaches to describe baryons are widespread. On the
one hand side there is the valence quark picture, which \eg enters
into nonrelativistic quark models \cite{Fey71}, bag models
\cite{Has78} or the description of baryons as quark-diquark bound
states \cite{Buc92}--\cite{Mey94}. On the other side the soliton
picture has successfully been applied to the investigations of baryon
properties \cite{Sky61}. The soliton description emerges from
large $N_c$-QCD studies \cite{Wit79a} and allows one to straightforwardly
incorporate the fruitful concept of chiral symmetry and its
spontaneous breaking. While the valence quark picture directly leads
to the quantum numbers of a physical baryon the soliton can only be
interpreted as a baryon after collective quantization. On the contrary
the soliton picture is certainly superior when considering features
which are related to chiral symmetry, \eg the matrix elements of the
axial singlet current \cite{Bro88}. These few examples
indicate that both approaches have only limited ranges of
applicability. Therefore a unification of the two approaches is
desirable. One such hybrid model is represented by the chiral bag
model \cite{Cho75}. It contains explicit quark
degrees of freedom inside the bag while a chiral meson cloud dwells
outside. According to the Chesire cat principle \cite{Nad85} physical
quantities should not depend on the radius of the bag. From a
conceptual point of view it would, however, be more appealing to have
the two approaches combined within one model with the preferred picture
selected by the dynamics. In this context the Nambu--Jona-Lasinio
(NJL-) model \cite{Nam61} is unique. Hence this model makes
possible a consistent unification by means of hadronization techniques
\cite{Rei90} without any double counting of correlations, because it
contains both, soliton solutions of meson fields
\cite{Rei88} (for a recent review see ref. \cite{Alk94}) as
well as diquark bound states \cite{Buc92,Ish93}.

Although the construction of the rigorous self-consistent solution
represents a complicated task we have been able to carry out the first step
towards this goal: We have solved the Bethe-Salpeter equation for a
diquark in the background of the chiral soliton. Furthermore we have
estimated the effects of the resulting diquark configuration on the
soliton. We will report on this progress in the present letter.

\section{\large \it Hadronized NJL-model}

As starting point we assume the NJL-model for two flavors
\be
 \L_{NJL} = \bar{q}\left(i\dslash-\hat{m}^0\right)q - \frac{1}{2} g
 j_{\mu}^a j^{\mu}_a
\ee
with a pointlike interaction of color octet flavor singlet currents
$j_{\mu}^a = \bar{q} \lambda_c^a \gamma_{\mu} q /2$. Here $q$ denotes
the quark spinors, $\hat{m}^0={\rm diag}(m_0^u,m_0^d)$ the current quark
matrix and $\lambda_c^a (a=1,..,8)$ are the generators of color
SU(3). The currents are rearranged by Fierz transforming into
attractive meson ($\bar{q}q$ color singlet) and diquark channels. The
latter represent $qq$ color anti-triplets and serve as
building blocks for baryonic color singlet states.

This quark theory has been converted into an effective theory of mesons
and baryons using path integral hadronization techniques \cite{Rei90}
\be
 \label{action}
 \A[\varphi, \Delta, \tilde{\Delta}] &=& \A_f[\varphi, \Delta,
 \tilde{\Delta}]  + \A_m[\varphi] + \A_{d}[\Delta, \tilde{\Delta}],
 \nonumber \\
 \A_f &=& \frac{1}{2} \TTr \log \GG^{-1}, \nonumber \\
 \A_m &=& -\frac{1}{8g}\int d^4x \left[\tr_{C,F,D} \varphi^2\right]
 ,\nonumber \\
 \A_d &=& -\frac{3}{8g}\int d^4x \left[\tr_{C,F,D} \tilde{\Delta}
 \Delta\right] .
\ee
For simplicity we have omitted the baryon fields and restricted
ourselves to the (pseudo-)scalar meson ($\varphi$) and scalar diquark
fields ($\Delta,\tilde{\Delta}$) in the isospin limit $m_0^u=m_0^d=:m_0$.
Because of the explicit color structure of the diquarks an additional
factor 3 appears in the mass term $\A_d$ for the diquarks as
compared to the mass term $\A_m$ of the mesons. $\A_f$ refers
to the fermion determinant with $\GG$ being the quark Greens function
\be
\label{green}
 \GG^{-1} = \left(\begin{array}{cc}
               \G^{-1}           & -\Delta C \\
               -C \tilde{\Delta} & -V\tilde{\G}^{-1}V^\dagger
            \end{array} \right) ,
\qquad \tilde{\G} = V^\dagger \G^T V ,
\ee
in the Nambu-Gorkov formalism of superconductivity \cite{Sch64}.
${\G}^{-1} = i \dslash - \varphi$ represents the normal quark
Greens function in the mesonic background $\varphi$ while $-\Delta C$ and
$-C \tilde{\Delta}$ are anomal Greens functions. The transformation
$V=JG$, which we have introduced for technical reasons, is a
combination of the self-adjoint unitary
transformation $J=i\beta\gamma_5$ and the G-parity operator
$G=e^{i\pi\tau_2}C$ with $C$ being the charge conjugation matrix in
the space of the Dirac matrices.

For the meson field $\varphi$ we make use of the polar decomposition
\be
 \varphi = \Phi\ U^{\gamma_5} = m\ U^{\gamma_5},
\ee
We have fixed the chiral radius $\Phi$ to its vacuum expectation
value and $U$ is the chiral field. The NJL model needs regularization
which introduces one more parameter, the cut--off $\Lambda$. We use
Schwinger's proper time description\cite{Sch51}. Altogether the
model contains three parameters: $m_0,g$ and $\Lambda$. As input
quantities we take the pion decay constant $f_\pi=93$MeV and the
pion mass $m_\pi=135$MeV leaving one undetermined parameter. This we
choose to be the constituent quark mass $m$, which is given in terms
of $m_0,g$ and $\Lambda$ as the solution of the Schwinger-Dyson equation.

\section{\large \it Bethe-Salpeter equation for diquarks in a solitonic
background}

In the following we are interested in the behavior of diquarks in the
background field of a soliton configuration of the meson fields. For
this purpose we adopt the well-known hedgehog ansatz
\be
\label{chiral-field}
 U = {\rm exp}\Big(i{\mbox{\boldmath $\tau$}}
 \cdot{\hat{\mbox {\boldmath $r $}} }\Theta(r)\Big)
\ee
with some given chiral angle $\Theta(r)$.

To derive the Bethe-Salpeter equation for the diquarks we expand the
effective action (\ref{action}) up to second order in the diquark field.
This allows us to evaluate the Nambu-Gorkov trace. The
remaining trace will be performed using the color-degenerated eigenstates
of the inverse quark propagator in the solitonic background
\be
\label{propagator}
{\G}^{-1}|\lambda\rangle = \left(i \dslash - \varphi\right)
|\lambda\rangle = \beta \left( i \partial_t - h_\Theta\right)
|\lambda\rangle = \beta \left( E - \epsilon_{\nu}\right) |E , \nu \rangle .
\ee
Note that the information about the soliton is contained in the quark
eigenstates $|\lambda\rangle$ and -energies $\epsilon_{\nu}$ of the
Dirac one-particle Hamiltonian
$h_\Theta=\mbox{\boldmath $\alpha$}\cdot \mbox{\boldmath $p$}
+\beta m U^{\gamma_5}$.

We consider an S-wave scalar diquark field. The only
ansatz compatible with the Pauli principle reads
\be
 \Delta(\mbox {\boldmath $r $},t) = \Delta_\alpha(r,t) \Gamma^\alpha
 \quad , \quad  \tilde{\Delta}(\mbox {\boldmath $r $}) =
 \Delta^*_\alpha(r,t) \Gamma^\alpha \quad  , \quad
 \Gamma^\alpha = - \frac{\lambda^a_C}{\sqrt{2}} \frac{\tau_2}{2} i
 \gamma_5
\ee
where $\lambda^a_C (a=2,5,7)$ are the antisymmetric Gell-Mann
matrices of the color group. These are the generators of a
$\bar{3}$--representation, which is equivalent to the spin one
representation of an SU(2) subgroup. The quadratic part of the
effective action for the diquarks may formally be expressed as
\be
 \label{diquarkaction}
 \A^{(2)}_{diq}\left[\Delta_\alpha , \Delta^*_\alpha\right] = \int
 \frac{d\omega}{2\pi} \left[ \int dr r^2 \int
 dr^\prime r^{\prime 2} \Delta^*_\alpha(r^\prime,-\omega)
 K(r^\prime,r;\omega) \Delta_\alpha(r,\omega) \right. \nonumber \\
 \left. - \frac{\pi m_\pi^2 f_\pi^2}{2 m_0m} \int dr r^2
 \Delta^*_\alpha(r^\prime,-\omega) \Delta_\alpha(r,\omega) \right] ,
\ee
where $\Delta_\alpha(r,\omega)$ denotes the Fourier-transform of the
$\Delta_\alpha(r,t)$, \etc. Furthermore we have made use of the
relation $g=6m_0m/m_\pi^2 f_\pi^2$. The bilocal kernel
$K(r^\prime,r;\omega)$ is represented via mode sums over the
eigenstates of $h_\Theta$ (\ref{propagator}). This kernel is local in
the frequency $\omega$ because the background field (\ref{chiral-field})
is static. The Bethe-Salpeter equation is
given as the equation of motion for the diquark fields
$\delta\A^{(2)}_{diq}/\delta\Delta^*_\alpha(r^\prime,-\omega)=0$:
\be
\label{BS}
 r^2\left[ \int  dr^\prime r^{\prime 2}  K(r^\prime,r;\omega)
 \Delta_\alpha(r^\prime,\omega)    - \frac{\pi m_\pi^2 f_\pi^2}{2m_0m}
 \Delta_\alpha(r,\omega) \right] = 0 .
\ee
For its solution we have adopted the procedure from the
treatment of meson fluctuations in the soliton background
\cite{Wei93b}. Since the quark spectrum is symmetric in the color
degrees of freedom the Bethe-Salpeter kernel $K(r^\prime,r;\omega)$ is
even in $\omega$. Therefore the solutions of (\ref{BS}) appear in
pairs $\pm \omega_{diq}$.

\vfill
\eject

\section{\large \it Numerical results and discussion}

\begin{figure}[t]
\caption{The dependence of the diquark energy $\omega_{diq}$ (solid
line) and the energy $2 \epsilon_{val}$ of two unbound quarks (long
dashed line) on the extension of the soliton. Also displayed is the
binding energy $E_B = 2 \epsilon_{val} -\omega_{diq}$ (short dashed
line). All results are for $m=450$MeV.}
\label{diqenerg}
\vspace{0.6cm}
\vskip0.5cm
\end{figure}

\begin{figure}[t]
\caption{Comparison between the pure soliton energy $E_{sol}$ (long
dashed line) and the diquark binding energy reduced soliton energy
$E_{sol}^{(diq)} = E_{sol}- E_B$ (solid line) for constituent
quark mass $m=450$MeV.}
\label{solenerg}
\vspace{0.6cm}
\vskip0.5cm
\end{figure}

In order to test the numerical treatment of
(\ref{diquarkaction},\ref{BS}) we have verified that the diquark
mass is reproduced in the absence of the soliton, \ie $\Theta=0$
\cite{Wei93a}.

As a test profile we have considered the self-consistent soliton
$\Theta_{\sc}(r)$ \cite{Alk94} of the NJL-model without
diquarks. In order to estimate the influence of the bound diquark
field onto the soliton we have introduced a dimensionless parameter
$a$, which measures the extension of the soliton profile, via
$\Theta(r)=\Theta_{\sc}(r/a)$. This choice is suggestive because the
energy $E_{sol}$ of the soliton (without diquarks) exhibits only a
moderate dependence on $a$. Hence this mode is expected to be very
sensitive on modifications of the model as \eg the incorporation of
diquarks. Obviously, all quantities depend on $a$ in a parametrical
way, in particular the binding energy of the diquarks, $E_B = 2
\epsilon_{val} -\omega_{diq}$, see figure \ref{diqenerg}. Here the
subscript $val$ refers to the eigenstate of $h_\Theta$ with the
smallest positive eigenvalue, \ie the valence quark state.

As a first non-trivial result we observe that $\omega_{diq} < 2
\epsilon_{val}$ even when the valence quark is strongly bound in the
soliton background. This especially implies that replacing two of the
three valence quarks in the soliton by a bound diquark
\underline{reduces} the total energy. This replacement just constitutes
the additive quark-diquark model for the baryon. In this additive
model the static energy of the soliton is given by $E_{sol}^{(diq)} =
E_{sea}+\omega_{diq}+\epsilon_{val} = E_{sol}- E_B < E_{sol} ( =
E_{sea}+ 3 \epsilon_{val}) $. Hence
we conclude that diquark correlations in the soliton background lower
the energy of the soliton. The dependence of $E_{sol}^{(diq)}$ on the
scaling parameter $a$ is shown in figure~\ref{solenerg}. Note that
$E_{sea}$ also depends on $a$, although only moderately. For a
constituent quark mass $m=450$MeV, which in the soliton picture
approximately reproduces the experimental $\Delta$-nucleon mass
difference \cite{Re89}, we observe a minimum for
$a \approx 0.94$. This indicates that the diquark correlations cause
some shrinking of the soliton profile.

\section{\large \it Conclusions and outlook}

In this letter we have reported on the first step towards a
description of baryons within a combined picture of a solitonic lump
and a bound system consisting of a quark and a diquark. For this
purpose we have considered the hadronized NJL model \cite{Rei90}, which
unambiguously contains both mesonic as well as diquark degrees of
freedom. We have solved the Bethe--Salpeter equation for the diquark
in the background of a soliton configuration. We have observed
that the appearance of a bound diquark is a genuine feature of the
model even when the valence quark is strongly bound by the soliton.
In an additive model for the baryon we have furthermore found that
the static energy is significantly reduced by diquark correlations.
Although the additive model merely represents an approximation this
result strongly indicates that in a more elaborated treatment the
incorporation of diquark correlations
will be important for the extension of the soliton picture of baryons.
On the other hand our estimates indicate that the feedback of the
bound diquark on the soliton is only moderate.

Here we have only considered a scalar diquark. Obviously it will
be equally interesting to see whether or not an axial diquark
is also bound by the soliton.

The treatment presented here offers a variety of possible extensions.
In the next step the quark--diquark interaction has to be taken into
account. This can be realized by solving (an approximation to) the
Faddeev equation \cite{Rei90} taking proper account of the Pauli
principle for the color degrees of freedom. This will directly provide
quantum numbers of physical baryons thus improving on the additive model.

Alternatively one might try to introduce the diquark field already
at the static level. This implies an {\it ansatz} for $\Delta_\alpha$
in the SU(2) color subgroup $\lambda_c^{(2,5,7)}$ entering the quark
Greens function (\ref{green}). This
aims at a self--consistent solution for both the colorless meson-- and
the colored diquark fields. As a consequence the quark sea will
additionally be polarized in the color degrees of freedom. The
associated dynamics will thus distinguish between diquark and anti--diquark
solutions. The resulting picture will be that of a multi--flavor --
multi--color U($N_F\times N_C$) soliton \cite{Kap91}.

\end{document}